\documentclass[12pt]{article}
\usepackage{epsf}

\usepackage{epsfig,graphics}
\usepackage {graphicx}
\usepackage {epsfig}
\usepackage {subfigure}
\usepackage {tabularx} 
\usepackage{rotate}	
\usepackage{slashed}

\usepackage{amsmath}
\usepackage{amsfonts}
\usepackage{amssymb}
\usepackage{graphicx}
\usepackage{cite}

\newcommand{\bmat}{\left(\begin{array}}
\newcommand{\emat}{\end{array}\right)}

\def\yzero{\smash{\hbox{$y\kern-4pt\raise1pt\hbox{${}^\circ$}$}}}

\def\beq{\begin{equation}}
\def\eeq{\end{equation}}
\def\beqa{\begin{eqnarray}}
\def\eeqa{\end{eqnarray}}

\def\-{\hphantom{-}}

\def\s2{\frac{1}{\sqrt2}}

\def\beq{\begin{equation}}
\def\eeq{\end{equation}}
\def\beqa{\begin{eqnarray}}
\def\eeqa{\end{eqnarray}}

\def\IF{\relax{\rm I\kern-.18em F}}
\def\II{\relax{\rm I\kern-.18em I}}
\def\IP{\relax{\rm I\kern-.18em P}}
\def\IC{\relax\hbox{\kern.25em$\inbar\kern-.3em{\rm C}$}}
\def\IR{\relax{\rm I\kern-.18em R}}

\def\Dsl{\,\raise.15ex\hbox{/}\mkern-13.5mu D} 
\def\IZ{Z\kern-.4em  Z}

\catcode`\@=11   
\newdimen\@rotdimen
\newbox\@rotbox  

\def\@vspec#1{\special{ps:#1}}
\def\@rotstart#1{\@vspec{gsave currentpoint currentpoint translate
   #1 neg exch neg exch translate}}
\def\@rotfinish{\@vspec{currentpoint grestore moveto}}
\def\@rotr#1{\@rotdimen=\ht#1\advance\@rotdimen by\dp#1%
   \hbox to\@rotdimen{\hskip\ht#1\vbox to\wd#1{\@rotstart{90 rotate}%
   \box#1\vss}\hss}\@rotfinish}
\def\@rotl#1{\@rotdimen=\ht#1\advance\@rotdimen by\dp#1%
   \hbox to\@rotdimen{\vbox to\wd#1{\vskip\wd#1\@rotstart{270 rotate}%
   \box#1\vss}\hss}\@rotfinish}%
\def\@rotu#1{\@rotdimen=\ht#1\advance\@rotdimen by\dp#1%
   \hbox to\wd#1{\hskip\wd#1\vbox to\@rotdimen{\vskip\@rotdimen
   \@rotstart{-1 dup scale}\box#1\vss}\hss}\@rotfinish}%
\def\@rotf#1{\hbox to\wd#1{\hskip\wd#1\@rotstart{-1 1 scale}%
   \box#1\hss}\@rotfinish}%
\def\rotate{\@ifnextchar[{\@rotate}{\@rotate[l]}}
\def\@rotate[#1]#2{\setbox\@rotbox=\hbox{#2}\@nameuse{@rot#1}\@rotbox}

\catcode`\@=12

\topmargin
-1.5cm
\textwidth
15.5cm
\textheight
23.5cm
\oddsidemargin
0.7cm
\evensidemargin
0.7cm

\begin{document}

\makeatletter
\@addtoreset{equation}{section}
\makeatother
\renewcommand{\theequation}{\thesection.\arabic{equation}}
\pagestyle{empty}
\vspace{-0.2cm}
\rightline{ IFT-UAM/CSIC-15-75}
\vspace{1.2cm}
\begin{center}


\LARGE{ Minkowski 3-forms, Flux String Vacua,  \\
Axion  Stability and  Naturalness} 
\\[13mm]
  \large{Sjoerd Bielleman,  Luis E. Ib\'a\~nez$  $, Irene  Valenzuela   \\[6mm]}
\small{
  Departamento de F\'{\i}sica Te\'orica
and Instituto de F\'{\i}sica Te\'orica UAM/CSIC,\\[-0.3em]
Universidad Aut\'onoma de Madrid,
Cantoblanco, 28049 Madrid, Spain 
\\[8mm]}
\small{\bf Abstract} \\[7mm]
\end{center}
\begin{center}
\begin{minipage}[h]{15.22cm}
We discuss the role of Minkowski 3-forms in flux string vacua.
In these  vacua all internal closed string  fluxes are in one to one correspondence with
quantized Minkowski  4-forms. 
By performing a dimensional reduction of the $D=10$ Type II  supergravity actions we find 
that the 4-forms  act as auxiliary fields of the Kahler and complex structure moduli
in the effective action.
We show that all the RR and NS axion dependence of the flux scalar
potential appears through the said 4-forms. Gauge invariance of these forms then 
severely restricts the structure of the axion scalar potentials. Combined with duality 
symmetries it suggests that all perturbative  corrections to the leading axion scalar
potential $V_0$  should appear as an expansion in  powers of $V_0$ itself. 
These facts  could have an
important effect e.g. on the inflaton models based on F-term axion monodromy. 
We also suggest that the involved multi-branched structure of string vacua 
provides  for a new way to maintain interacting scalar masses  stable against 
perturbative corrections.

\end{minipage}
\end{center}
\newpage
\setcounter{page}{1}
\pagestyle{plain}
\renewcommand{\thefootnote}{\arabic{footnote}}
\setcounter{footnote}{0}



\tableofcontents

\section{Introduction}

Consistency of Poincare invariant field theory implies that the possibilities for Lorentz structure of
massless fields is quite limited. Fermions must have spin 1/2 or 3/2, whereas bosons 
should have a Lorentz structure of any of the kinds $C^0,C^\mu,C^{\mu\nu}$  or $g^{\mu\nu}$,
with $g$ the graviton and the C-fields being antisymmetric. This list  should be extended to
include  3-index antisymmetric tensors  $C^{\mu \nu \rho}$. At first sight this extra possibility looks irrelevant, since 
a Minkowski 3-form has no propagating degrees of freedom. However the presence of such fields may lead to important
physical implications. A very recent example of this fact is discussed in \cite{dixon}, in which it is shown that
the ultraviolet behaviour  of pure gravity amplitudes changes if 3-form contributions are included in loops, in spite of
not having propagating degrees of freedom.  More well known is the fact that 
the corresponding  field strength $F^{\mu \nu \rho \sigma}$ may be non-vanishing and permeate  space-time giving rise
to a constant contribution to the cosmological constant, and hence to new (quantized) degrees of freedom. 
  Due to this fact Minkowski 4-forms have been considered 
in the past in trying to address the cosmological constant problem \cite{duff1,hawking,duff2,wu,duncan}.  More specifically
Brown and Teitelboim \cite{BT}  considered a background 4-form  field strength in space-time, contributing to
the vacuum energy.
Membranes coupling to the $C^{(3)}$ form can nucleate and give rise to jumps in the c.c. They suggested this contribution adjusts  itself dynamically 
to cancel the rest of the contributions to the c.c.  This 4-form is assumed not to couple directly to other 
fields in the theory. The main  difficulty with this approach is that the 4-form steps required to cancel the c.c. should be extremely tiny and is
difficult to construct a working model with the required properties.

Bousso and Polchinski \cite{BP}  suggested also to consider the contribution of 4-forms to the c.c. within the context of
string theory (see also \cite{MR}).  They argued  that in string theory plenty of Minkowski 4-forms appear upon compactification and 
that their values are quantized. There is then 
a {\it discretum}   in which the  individual  (large) 4-form values could conspire to yield a detailed
(almost) cancellation of the c.c. if the number of 4-forms and their possible quantized values is sufficiently large.
The structure of the scalar potential has the schematic form
\beq
V\ =\ \sum_{i} \  F_i^2 \ -\ V_{obs}
\eeq
where $F_i = \epsilon^{\mu \nu \rho \sigma}F^i_{\mu \nu \rho \sigma}$ and $-V_{obs}$ denotes the remaining
contributions, typically yielding a negative value.
In this case the cancellation is not dynamical but is assumed to occur on the basis of anthropic arguments.
A difficulty  with this proposal so formulated  is that within string vacua one cannot separate the issue of the c.c. from that of 
moduli fixing and one expects the 4-forms to couple to the moduli, making the situation far more complicated.
As is well known, soon after a general approach to fix all moduli within Type IIB string theory
vacua was proposed \cite{KKLT}, in which {\it internal} RR and NS fluxes are turned on  \cite{GKP} to fix
the complex structure moduli and dilaton in Type IIB orientifolds, with the Kahler moduli assumed to be fixed
by non-perturbative effects.  Since then a large amount of effort has been dedicated to the issue of moduli fixing,
involving internal fluxes \cite{LVS,fluxes}. Still the possible role  of Minkowski 4-forms has been rarely discussed.

Minkowski 4-forms were discussed in papers by  Dvali  \cite{Dvali1}  in which it was shown  that the 
usual strong CP problem and its axion solution may be elegantly described in terms of a
{\it composite} 3-form, the QCD Chern-Simons term $C^{(3)}$, with a dynamical 4-form proportional to 
$F\wedge F$. Here the PQ solution to the strong CP problem corresponds to the 3-form becoming massive 
via a coupling to a 2-form $B_{\mu \nu }$, the latter being the dual of a standard axion. 

More recently  Kaloper and Sorbo  \cite{KS,KLS} showed that 4-forms in field theory provide for a natural
definition of  quadratic chaotic inflation \cite{chaotic}, stable under large field trips of the inflaton. Schematicaly, one starts from an
action including not only a quadratic piece for the 4-form but a coupling to an axion-like field $\phi$
\beq
{\cal L} \ =\ -\ F_4^2 \ +\ \mu \phi F_4 \ +\ \dots
\eeq
with $\mu$ some mass parameter. Impossing $dC_{3}=F_4$ through a Lagrange multiplier $q$ and
upon using the equations of motion for $F_4$,  one finds a quadratic scalar potential of the form
\beq
V_{0}\ =\ \frac {1}{2} \left( q\ +\ \mu\phi\right)^2
\eeq
where $q$ is interpreted as a $F_4$ vev. Membranes couple to the 3-form $C_3$ and induce 
changes $\Delta q= e$, where $e$ is the membrane charge. The interesting point is that this is
{\it not} a scalar potential but rather a {\it family} of potentials or different branches parametrized by the value of $q$.
The family of potentials has a discrete shift  symmetry 
\beq
\phi \rightarrow \phi \ + \ \phi_0 \ ,\   q\rightarrow q-\mu \phi_0
\eeq
which is spontaneously broken when a minimum $\phi=-q/\mu$ is chosen. For each local minimum we have
a quadratic potential, which can be used e.g. to induce chaotic inflation if $\phi$ is identified with the inflaton.
The above description in terms of a 4-form is a way of gauging a shift global symmetry for a scalar field without introducing new degrees of freedom.
One can formulate the same system by using the dual 2-form $B_2$ instead of the scalar $\phi$. 
Here, as in \cite{Dvali1} , $C_3$ gets massive by combining with $B_2$, yielding a massive degree of freedom.
One then obtains  an action of
the schematic form \cite{Dvali1,msu}
\beq
{\cal L} \ = \ -F_4^2 \ -\ \frac {\mu^2}{2} |dB_2\ -\ C_3|^2 \ +\dots
\eeq
This action is obviously invariant under a gauge transformation
\beq
 B_2\rightarrow B_2 + \Lambda_2 \ \ ,\ \  C_3 \rightarrow C_3 +d\Lambda_2
\eeq
which corresponds to the above shift symmetry. 
This shift symmetry is expected to be broken in a complete theory by non-perturbative effects.
However, 
what makes this elaborated  construction of
a simple quadratic potential interesting is that the symmetries will protect the potential from perturbative and
Planck suppressed  corrections.
Indeed, gauge invariance of $F_4$ and the shift symmetry of $\phi$ 
force the corrections to appear in powers $(F_4^2/M^4)^n$, with $M$ the ultraviolet cut-off of the
theory, rather than arbitrary powers of $\phi$. Thus corrections to an inflationary potential should appear as powers of $V_0/M_p^4$, 
which will be very small for an inflaton potential $V_0<(10^{16}GeV)^4$. This is crucial to get 
stability of large field inflation in these schemes.

This Kaloper-Sorbo Lagrangian is a 4D field theory avatar of a somewhat analogous structure 
found in the {\it monodromy inflation} models of  \cite{mono,msu,more mono,higgsotic}. In those models large field inflation
is attained by coupling an axion-like  periodic field to an external source of energy, like e.g. a brane tension.
Upon a period the field gets a shift in energy,  so that the field does not come to the same point but
rather  perform a large trans-Planckian  excursion.  In the recent paper of Marchesano, Shiu and Uranga \cite{msu}
it has been explicitly shown how a structure analogous  to that of the KS Lagrangian appears in
specific string constructions.

 In the present paper we study in a systematic way the role of  Minkowski 4-forms in Type II,  $D=4$, $N=1$ 
 orientifold vacua and discuss  to what extent the above discussed 4-form avatars do appear in
 compactified  string theory.  We also study the connection between the  internal RR and NS fluxes 
 abundantly used in moduli fixing and the Minkowski 4-forms.  We analyse in more detail the case of
 Type IIA  orientifold $N=1$  flux vacua, in which the discussion is more transparent,  but also present
 analogous  results for the Type IIB case. 
 In the former case some of the conclusions are as follows
 
 \begin{itemize}
 
 \item    RR and NS closed string fluxes through internal cycles are in one to one correspondence to
 Minkowski 4-forms. 
  These  4-forms act as auxiliary fields of both Kahler and complex structure moduli as well as for the
 $N=1$ supergravity multiplet.

 \item  The full dependence of the flux scalar potential on  RR and NS axions goes always through combinations
 of Minkowski 4-forms.  As a result the scalar potentials  of string flux vacua are not any random sugra potential but
 have a branched structure. The potential has the general form
 \beq
 V_{4-forms}\ =\ \sum_i f_{ij}(ReM_a)\  F_4^iF_4^j \ +\sum_i  F_4^i\ \Theta_i(ReM_a, ImM_a)\ +\  V_{local}(ReM_a) \ .
 \label{estructura}
 \eeq
 Here $M_a$ denote collectively both  Kahler and complex structure moduli, and $ImM_a$ denote the RR and NS axions.
 The functions  $\Theta_i$ come from the Type IIA Chern-Simons couplings and contain  polynomials of  the axion fields
 with coefficients involving linearly  the internal fluxes.
 $V_{local}$ contains the contribution of the D-branes and orientifold planes to the potential, which can be re-expressed 
 in terms of the $ReM_a$ upon imposing RR tadpole cancellation. Upon applying the equations of motion for the 4-forms
 the full scalar potential is obtained, with an axion dependent part which is always positive definite.  
  
 \item  The above scalar  potential is in some sense a string multi 4-form and multi-flux generalisation of the Kaloper-Sorbo structure
 in which the quadratic potential is replaced by more general (up to order six) polynomials.  The role of the shift symmetry is
 played by the duality symmetries of the compactified theories. Under  $R\leftrightarrow 1/R$ duality symmetries the different 
 Minkowski 4-forms  transform into each other. As in the KS field theory model, gauge invariance of the 4-forms
 combined with the duality symmetries of the compactification constrain the corrections to the potential to come
 suppressed by powers of $V_0/M_p^4$. This shows that flux string vacua is a natural arena to construct large
 field inflaton models with a stable potential.
  
 \end{itemize}

 The structure in eq.(\ref{estructura}) resembles the one discussed by Beasley and Witten in the context of M-theory compactified in $G_2$ manifolds $X$
 in the presence of $G_4$ flux \cite{BW}. They  found that, although the superpotential $W$  depends explicitly only on the $G_4$ flux supported on $X$, 
 it also describes the breaking of SUSY by $G_4$ flux in Minkowski. The resulting scalar potential is also branched, in analogy with the
 Schwinger model in two dimensions \cite{Coleman}.

 This structure leads to  families of scalar potentials parametrized by specific flux choices, some  of which are related by 
 orbits of duality transformations.  As expected, there can be transitions from one potential to another by membrane nucleation.
 This has been analysed in a context similar to ours in \cite{domainwalls}.
 The membranes in Type IIA come from D2, D4, D6 and D8-branes wrapping even cycles (for RR 4-forms) and NS5 branes wrapping
 3-cycles (for NS 4-forms).
  Analogous conclusions  hold for Type IIA vacua with geometric fluxes. In this case  the nucleating membranes
 will be KK5-branes wrapping  3-cycles. 
 A similar story also applies to  $N=1$ Type IIB orientifolds with RR and NS fluxes, which we describe more briefly. 
 We also briefly touch upon the issue of non-geometric fluxes.
 In the Type IIB  case the natural objects which appear are complex 4-forms, involving  the complex dilaton as well as 
 both RR and NS fluxes in their definition.
 
 We also suggest that the above structure of symmetries may provide for a new way to obtain an interacting 
 theory of scalars in which stability against loop corrections may be obtained.  This would be a consequence of the
 multi-branched structure of the axion scalar fields yielding a corrected potential which is itself an expansion in powers of
 the uncorrected potential. We also speculate about possible applications of this idea.

 The structure of this paper is as follows. In the next section we recall a few facts about Minkowski 4-forms in general. In section 3 we study the
 structure of Minkowski 4-forms in Type IIA orientifolds with RR and NS fluxes. We perform the dimensional reduction starting from the
 $D=10$ Type IIA action and focus on the couplings of the Minkowski 4-forms. We show how they behave as moduli auxiliary fields
 and how they are invariant under a class of discrete symmetries involving both RR and NS axion shifts as well as internal flux transformations.
 We also discuss in the toroidal case the action of $R\leftrightarrow 1/R$ dualities  as well as how  the introduction of geometric fluxes
 modifies the setting.  In section 4 we address the case of Type IIB orientifolds and how in this case the RR and NS 4-forms combine to yield 
 complex auxiliary fields, but a structure otherwise analogous to that of the Type IIA case.  We also discuss briefly how 4-forms may  arise 
 from the open string sector, by dimensionally reducing the duals of the $F_2$ gauge field strengths, and discuss in some detail the 
 example of reference \cite{higgsotic}. In section 5 we present a general discussion of implications of the uncovered symmetry structure 
 for the stability of scalar potentials against perturbatione corrections. We briefly discuss the case of inflation and a possible new way to obtain 
 naturally light interacting scalars. Some conclusions are left for section 6.

\section{Minkowski 3-forms}

Before turning to Type II orientifold compactifications, let us recall a few facts about 3-forms (see e.g. \cite{BP,MR,Dvali1,KS,Dudas,QT}).
The bosonic  action of a 3-form includes terms
\beq
S\ =\  -\int d^4x  \sqrt{-g} \frac{1}{48} F_{\mu\nu\rho\sigma}F^{\mu \nu \rho\sigma} \ +\  S_{bound} \ +\ S_{mem} \ .
\eeq
Here $S_{bound}$ includes some boundary terms which do not modify the equations of motion and will not play 
a role in our discussion, so will not be displayed here. On the other hand $S_{mem}$ describes the possible coupling 
of $C_3$ to membranes, i.e.
\beq
S_{mem}\ =\ q\int_{D_3} d^3\xi \epsilon^{abc}\ C_{\mu\nu\rho}\left(
\frac {\partial X^\mu}{\partial\xi^a}\frac {\partial X^\nu}{\partial\xi^b}\frac {\partial X^\rho}{\partial\xi^c}\right) \ ,
\eeq
where the membrane charge $q$ has dimensions of mass$^2$ and $D_3$ is the membrane world volume.
Away from the membranes the equations of motion for $C_3$ force $F_4$ to be constant, i.e.
\beq
F_{\mu\nu\rho\sigma}\ =\ f\epsilon_{\mu\nu\rho\sigma}
\eeq
where $f$ is a constant. 
 In the presence of membrane domain walls, the value of this constant varies 
as $\Delta f= q$ as one goes across the wall.  As argued e.g. in \cite{BP}  the value of the 4-form  in string theory is quantized in
units of the membrane charge, i.e.
\beq
f\ =\ nq \ \ ,\ \ n\in {\bf Z} \ .
\eeq
In the case of generic string compactifications we will have multiple 4-forms, some coming directly from dimensional 
reduction and others upon expanding higher order antisymmetric RR or NS tensors in harmonics in the compact directions.
In addition, as we will see, unlike the BT or BP scenarios, the 4-forms have couplings to the axions and moduli of the 
compactification, with a structure for each 4-form
\beq
F^2\ +\ F \theta(\phi_i) \ ,\  F=F_{\mu \nu \rho \sigma}\epsilon^{\mu \nu \rho\sigma}
\eeq
with $\theta$ a  function of the axions and moduli. Upon integration by parts the second piece may be written as
\beq
C_{\nu\rho\sigma}\ J^{\nu\rho\sigma}  \ \ ;\ \ J^{\nu\rho\sigma}=\epsilon^{\mu\nu\rho\sigma}\partial_{\mu} \theta(\phi_i) 
\eeq
This current is conserved, i.e., $\partial_\nu J^{\nu\rho\sigma}=0$ and the action is invariant 
under the gauge transformations.
\beq
C_{\nu\rho\sigma} \ \longrightarrow \  C_{\nu\rho\sigma}\ +\   \partial_{[\nu}\Omega_{\rho\sigma]} \ .
\eeq 
%

For 4-forms  in which  $\theta(\phi_i)$ is just a  linear function of   a RR or NS  axion field, the structure 
of its contribution  to the action is analogous to a Kaloper-Sorbo action. 
In this case one can dualise the
axion into a Minkowski 2-form in the usual way, with
\beq 
\partial_\mu \phi \ =\  \epsilon_{\mu\nu\rho\sigma} \partial^\nu B^{\rho\sigma} \ .
\eeq
Then the $\phi F_4$ coupling becomes
\beq
C_{\nu\rho\sigma}(\partial^\nu B^{\rho\sigma})
\label{2Higgs}
\eeq
indicating how through a Higgs mechanism the 3-form gains a gauge invariant mass by swallowing the 2-form. This is the dual of
the axion becoming massive in the KS setting, and is what Dvali used for his reinterpretation of the QCD axion
physics \cite{Dvali1}. The 3-form and 2-form have then gauge transformations
\beq 
C_{\nu\rho\sigma}\rightarrow C_{\nu\rho\sigma}+\partial_{[\nu}\Omega_{\rho\sigma]} \ ;\  B_{\rho\sigma}\rightarrow B_{\rho\sigma} +\Omega_{\rho\sigma} \ .
\eeq
This leads to a massive 3-form multiplet, which now contains a massive scalar
degree of freedom. This structure of a massive scalar may be connected also with torsion cycles in string compactifications, as
emphasized in \cite{msu}.

Massless 3-forms may be embedded into $N=1$ supersymmetric multiplets. They naturally appear as auxiliary fields in non-minimal
versions of the $N=1$  chiral multiplet  \cite{gates1,gates2,binetruy,ovrut,binetruy2,girardi2,louis3form,louisdavid,Deo,Bandos,Nishino}.
And essentially correspond to replacing 
one or both of the real auxiliary fields of a chiral multiplet by corresponding 4-forms. Similarly, one can formulate non-minimal 
$N=1$ sugra multiplets with one or two real scalar auxiliary fields being replaced by 4-forms.  Still these type of multiplets have not been
discussed much in the literature. In \cite{louis3form} the SUSY action of a non-minimal chiral multiplet $S$ including one 4-form auxiliary field
is discussed in detail. The corresponding superfield may be defined as
\beq
S \ =\ -\frac {1}{4} {\overline D}^2V \ ,
\eeq
where $V$ is a real multiplet with the same content as a standard vector multiplet, but with the vector field 
replaced by $\epsilon_{\mu\nu\rho\sigma}C^{\nu\rho\sigma}$. The chiral $S$ field has then an expansion
\beq
S\ =\ M\ +\ i\theta\sigma^\mu{\overline \theta }\partial_\mu M\ +\ \frac {1}{4} \theta \theta {\overline \theta }{\overline \theta }\square M\ +\ \sqrt{2}\theta \lambda 
\ +\ \frac {i}{\sqrt{2}} \theta \theta {\overline \theta}{\overline \sigma }^\mu \partial_\mu \lambda \ +\ \theta \theta (D\ +\ iF)  \ ,
\eeq
with $F=\epsilon_{\mu\nu\rho\sigma}F^{\mu\nu\rho\sigma}$ and $D$ an auxiliary  real scalar.  This multiplet contains on-shell one complex scalar $M$ and one Weyl fermion $\lambda$.
It can combine with a linear supermultiplet $L$, which includes a 2-form antisymmetric field $B_2$,  to yield a massive 3-form multiplet. This is a SUSY generalisation of the
Higgs mechanism described around eq.(\ref{2Higgs}).  In addition these non-minimal chiral super fields $S$ can have superpotential couplings in superspace, i.e.
\beqa
S_W\ &=& \  \int d^2\theta d^2{\overline\theta}\  S^a{\overline S}^a \ +\ \int d^2\ \theta W(S) \ +\ \int d^2{\overline \theta }\ W^*({\overline S}) \  =\\ 
&-& |\partial M|^2 + D^aD^a+F^aF^a+W_a(D^a+iF^a)+W^*_a(D^a-iF^a)+\  ...
\eeqa
where $W_a$ denotes derivative with respect to $S_a$.
Using the equations of motion for $C_3$ one gets $F^a=Im(W_a)+f_a$, with $f_a$ a constant. Then the scalar potential has the form
\cite{binetruy,louis3form,louisdavid}
\beq 
V_S\ =\ |W_a\ +\ if_a|^2 \ .
\label{shift1}
\eeq
This agrees with the result obtained for standard chiral multiplets with the replacement $W_a\rightarrow W_a+if_a$.  
Let us advance that this multiplet is not enough to describe the structure of 4-forms that we find in Type IIA and IIB orientifolds.
In particular we find that  for the Kahler(complex structure) moduli in IIA(IIB) orientifolds both auxiliary fields of a chiral multiplet
are replaced by 4-forms.

\section{4-forms in Type IIA orientifolds }

We turn now to describe how 4-forms appear in Type IIA orientifold compactifications down to four dimensions. The compactification of ten-dimensional massive Type IIA string theory on a Calabi-Yau threefold in the presence of background fluxes has been thoroughly studied in e.g. \cite{LouisGrimm,LouisMicu,Villadoro,dewolfe,Luis2}. Here we perform the same compactification but keeping trace of all the Minkowski 4-forms which appear upon dimensionally reducing the 10d RR and NSNS fields. This leads to a new formulation of the scalar potential in terms of Minkowski 4-forms as in eq.\eqref{estructura} and the intriguing result that the full dependence of the flux scalar potential on RR and NS axions comes only through couplings to the said 4-forms.

\subsection{4-forms , RR and NS fluxes in IIA orientifolds}

Let us consider Type IIA string theory compactified on a Calabi-Yau threefold $Y$ in the presence of O6 planes. The massless ten-dimensional bosonic content of the closed string spectrum contains the metric, the dilaton and the antisymmetric two-form $B_2$ from the NS-NS sector and the p-form fields $C_p$ from the RR sector. We will work in the democratic formulation \cite{ortin}  in which all the p-form fields $C_p$ with $p=1,3,5,7$ are present, so we will have to impose the Hodge duality relations
\beq
G_6=-*_{10}G_4\quad ,\quad G_8=*_{10}G_2\quad ,\quad G_{10}=-*_{10}G_0
\label{Pdualities}
\eeq
at the level of the equations of motion in order to avoid overcounting of the physical degrees of freedom. The gauge invariant field strengths are defined as \cite{Villadoro,ortin}
\beq
G_p=dC_{p-1}-H_3\wedge C_{p-3}+\mathcal{F}e^B
\label{Gp}
\eeq
where $H_3=dB_2$, $F_{p}=dC_{p-1}$ and $\mathcal{F}$ is a formal sum over all the RR fluxes $F_{p}$. The background field strength $G_0$ may be regarded as the mass parameter (also known as Romans mass) of massive Type IIA supergravity, $G_0=-m$.
The massless 4d fields (before introducing the fluxes) are in one-to-one correspondence with the
harmonic forms of the internal manifold Y, so the multiplicity is counted by the dimension of the cohomology groups $H^{(p,q)}(Y )$.
 To implement the orientifold projection we split the harmonic forms into forms with even or odd parity under the orientifold projection. The elements of the cohomology basis satisfy the following relations,
\beqa
\int_Y \omega_\alpha \wedge \tilde \omega^\beta=\delta_\alpha^\beta \ ,&\quad \alpha,\beta\in \{1\dots h_+^{(1,1)}\}\\
\int_Y \omega_a \wedge \tilde \omega^b=\delta_a^b \ ,&\quad a,b\in \{1\dots h_-^{(1,1)}\}\\
\int_Y \alpha_K \wedge \beta^L=\delta_K^L \ ,&\quad K,L\in \{1\dots h^{(2,1)}+1\}
\eeqa
where $\omega,\tilde\omega,\alpha$ denote a 2-form, 4-form and 3-form respectively. Notice that since the volume form is odd under the orientifold projection and the Hodge star involves contraction with the volume form, the dual form of an odd 2-form $\omega_a$ is actually an even 4-form $\tilde\omega_a$. Therefore $\omega_a\in H^{(1,1)}_-$ and $\tilde \omega_a\in H^{(2,2)}_+$ while $\omega_\alpha\in H^{(1,1)}_+$ and $\tilde \omega_\alpha\in H^{(2,2)}_-$. Analogously $\alpha_K\in H^{3}_+$ and $\beta_K\in H^{3}_-$. The metric, the dilaton, $C_3$ and $C_7$ are even under the orientifold projection while $B_2$, $C_1$ and $C_5$ are odd.

We are interested in the presence of Minkowski 3-form fields in the fluxed induced scalar potential. In addition to the universal RR 3-form $C_3$ one can also get 3-forms by dimensionally reducing higher RR and NSNS fields, $C_5$, $C_7$, $C_9$ and $H_7$, and considering three of the indices in Minkowski space. By allowing also for the presence of internal fluxes, the RR field strengths can be expanded as
\beqa
\label{fluxes}
\nonumber F_0=-m\ ,\quad F_2=\sum_iq_i\omega_i\ ,\quad F_4=F_4^0+\sum_ie_i\tilde\omega_i\\ F_6=\sum_iF_4^i\omega_i+e_0dvol_6\ , \quad F_8=\sum_aF_4^a\tilde \omega_a\ ,\quad F_{10}=F_4^mdvol_6
\label{fluxes}
\eeqa
where $i,a=1,\dots,h_-^{(1,1)}$. The parameters $e_0,e_i,q_i,m$ refer to internal RR fluxes on $Y$ and we get $2h_-^{(1,1)}+2$ Minkowski 4-forms labelled by $F_4^0$, $F_4^i$, $F_4^a$ and $F_4^m$. 
Similarly the NS $H_3$ background is intrinsically odd under the orientifold projection so it can be expanded as
\beq
H_3=\sum_{I=0}^{h_{2,1}^-} h_I\beta_I
\eeq
while the dual $H_7$ can be expanded in terms of even 3-forms
\beq
H_7=\sum_I H_4^I\wedge \alpha_I
\eeq
obtaining $h_{2,1}^++1$ additional Minkowski 4-forms $H_4^I$ coming from the NSNS sector.
Moreover, the fields $B_2$ and $C_3$ can be expanded as
\beq
B_2=\sum_ib_i\omega_i\ ,\quad C_3=\sum_Ic_3^I\alpha_I
\label{B2C3}
\eeq
where $b_i$ and $c_3^I$ are 4d scalars and correspond to the axionic part of the complex supergravity fields $T,S,U$ as follows,
\beqa
ImT_i=-\int B_2\wedge \tilde\omega^i=-b^i   \ ;    \ \  i=1,..,h_-^{11} \\
ImU_i=\int C_3\wedge \beta^i=c_3^i  \ ;     \ \  i=1,..,h_+^3  \\
ImS=-\int C_3\wedge \beta^0=-c_3^0   \ .
\label{TSU}
\eeqa
The Hodge  dualities of eqs.\eqref{Pdualities} relate the Minkowski 4-forms and the internal magnetic fluxes as we proceed to explain in the following. Separating each  field strength into Minkowski and internal parts and using eq.\eqref{fluxes}, the duality relations given by \eqref{Pdualities} imply
\beqa
*_4F_4^0=\frac1k(e_0+e_ib^i+\frac12 k_{ijk}q^ib^jb^k-\frac{m}{3!}k_{ijk}b^ib^jb^k-h_0c_3^0- h_ic_3^i)\nonumber\\
*_4F_4^i=\frac{g^{ij}}{4k}(e_j+k_{ijk}b^jq^k-\frac{m}{2}k_{ijk}b^jb^k)\nonumber\\
*_4F_4^a=4kg^{ab}(q_b-mb_b)\nonumber\\
*_4F_4^m=-m\label{F4m}
\eeqa
where $g_{ij}=\frac{1}{4k}\int \omega_i\wedge * \omega_j$ is the metric in the Kahler moduli space, $k$ is the volume and $k_{ijk}$ the topological triple intersection number.

The type IIA ten dimensional supergravity action can be divided into three terms,
\beq
S_{IIA}=S_{RR}+S_{NS}+S_{loc}
\eeq
where the RR and NSNS actions  are  given by
\beq
S_{RR}=-\frac{1}{8k_{10}^2}\int_{R^{1,3}\times Y}\sum_{p=0,2,4,6,8,10}G_p\wedge *_{10}G_p+\dots\ ,\quad  S_{NS}=-\frac{1}{4k_{10}^2}\int_{R^{1,3}\times Y}e^{-2\phi}H_3\wedge *_{10}H_3
\eeq
and $S_{loc}$ refers to the contribution from localized sources like D6-branes and O6-planes. Let us start analyzing the part of the action involving the RR fields. 
By using the duality relations \eqref{Pdualities}, the kinetic terms for the RR fields can be written as
\beq
-\frac{1}{2}\sum_{p=0,2,4,6,8,10}G_p\wedge *_{10}G_p=G_4\wedge G_6+G_2\wedge G_8+G_0\wedge G_{10}
\eeq
Plugging eqs.\eqref{fluxes}-\eqref{B2C3} into the above RR action and integrating over the internal dimensions 
we get the following effective scalar potential in four dimensions
\begin{multline}
V_{RR}=-\frac{1}{2}\left[F_4^0\left(e_0+ b^ie_i+\frac12 k_{i jk}b^ib^jq_k-\frac{m}{6}k_{ijk}b^ib^jb^k\right)+\right.\\
\left.+F_4^i\left(e_i+k_{ijk}b^jq_k-\frac12 mk_{ijk}b^jb^k\right)+F_4^a(q_a-mb_a)-kmF_4^m\right]
\end{multline}
This scalar potential can be rewritten by using eq.\eqref{F4m} in the general form
\begin{multline}
V_{RR}=-\frac{1}{2}\left[-k F_4^0\wedge *F_4^0+2F_4^0\rho_0-4kg_{ij}*F_4^i\wedge F_4^j+2F_4^i\rho_i-\right.\\
\left.-\frac{1}{4k}g_{ab}F_4^a\wedge *F_4^b+2F_4^a\rho_a+kF_4^m\wedge *F_4^m\right]
\label{VRR}
\end{multline}
already discussed in the introduction, in which the relations \eqref{F4m} arise as equations of motion for the 3-forms. Since $^{**}F_4=-F_4$ the contribution to the potential energy is positive.
Notice that although the Minkowski 3-forms have no dynamical degrees of freedom in four dimensions, the kinetic terms of these 3-forms lead to a Minkowski background which also contributes to the scalar potential of the theory. In addition we have some Chern-Simons couplings of the Minkowski 4-forms to the functions
\beqa
\rho_0=e_0+ b^ie_i+k_{i jk}\frac12 q_ib^jb^k-\frac{m}{6}k_{i jk}b^ib^jb^k-h_0c_3^0- h_ic_3^i\nonumber\\
\rho_i=e_j+k_{ jkl}b^kq^l-\frac{m}{2}k_{jkl}b^kb^l\nonumber\\
\rho_a=q_b-mb_b\nonumber\\
\rho_m=-m
\label{rho}
\eeqa
depending polinomially on the axionic fields and the internal fluxes. Analogously, the kinetic term for the NSNS field leads to the following contribution,
\beq
V_{NS}=\frac12 e^{-2\phi} c_{IJ}H_4^IH_4^J
\eeq
where $c_{IJ}=\int \beta_I\wedge * \beta_J$ is the metric on the complex structure moduli space. 
By Hodge  duality the Minkowski 4-form background is related to the NS internal flux by
\beq
*H_4^I=h^I\label{H40}
\eeq
The contribution from the localized sources can be written as \cite{Villadoro}
\beq
V_{loc}=\sum_a \int_\Sigma T_a \sqrt{-g}\ e^{-\phi}
\eeq
where $T_a$ is the tension of the object and $\Sigma$ the worldvolume.
Assuming that tadpole cancellation is satisfied, this contribution can be related to the fluxes and the real part of the moduli so that \cite{Villadoro}
\beq
V_{loc}=\frac12 e^K v_iv_jv_k k_{ijk}(mh_0 s-mh_i u^i) \ ,
\eeq
with $s,u_i,v_i$ the real parts of the $S,U_i,T_i$ moduli respectively.
which is independent of the configuration of localized sources as long as they preserve $N=1$ supersymmetry.
Combining all the pieces and using \eqref{F4m} we get the following scalar potential
\beq
V=\frac{k}{2}|F_4^0|^2+2k\sum_{ij}g_{ij}F_4^iF_4^j+\frac{1}{8k}\sum_{ab}g_{ab}F_4^aF_4^b+k|F_4^m|^2+\frac{1}{2s^2}\sum_{IJ}c_{IJ}H_4^IH_4^J+V_{loc}
\label{VIIA}
\eeq
which in terms of the moduli and the internal fluxes becomes
\begin{multline}
V=\frac{1}{2k}(e_0+e_ib^i+\frac12 q_ik_{ijk}b^jb^k-\frac16 mk_{ijk}b^ib^jb^k)^2+\\
+\frac{g^{i\bar j}}{8k}(e_i+q^kk_{ikl}b^l-\frac12 mk_{ikl}b^kb^l)(e_j+q^mk_{jmn}b^n-\frac12 mk_{jmn}b^mb^n)+\\
+2kg_{ij}(q^i-mb^i)(q^j-mb^j)+km^2+\frac{1}{2s^2}\sum_{IJ}c_{IJ}h^Ih^J+V_{loc}
\label{LouisMicu}
\end{multline}
as has been previously obtained in the literature \cite{LouisMicu}. This potential can also be recovered from the standard Cremmer et al. supergravity description in terms of the $N=1$ 4d effective Kahler potential and superpotential, see \cite{LouisGrimm}.

We would like to recall that the full axionic part of the scalar potential can be written in terms of the above couplings to Minkowski 3-form fields and it is always positive definite. This is one of the main results of the paper.

It is worth mentioning a subtlety regarding the process of integrating out the 3-form fields. By looking at \eqref{VRR} the equation of motion for the 3-form field implies
\beq
d(*_4F_4-\rho)=0\rightarrow *_4F_4-\rho=c
\eeq
where $c$ is a constant and $\rho$ the function depending on the axionic moduli defined in \eqref{rho}. This would imply a shift on the 4-form background leading to a priori new terms in the scalar potential that can not be recovered from the standard Cremmer et al. supergravity description.
In particular the shifts would appears as quantized spurion insertions
which could have important  implications for moduli fixing and the search of de Sitter vacua. 
These  shifts agree with the results of \cite{binetruy,louis3form,louisdavid}  for which a 4-form acting as an auxiliary field implies a shift on the scalar potential with respect to the standard supergravity formula. While valid from a pure effective 4d approach, our 4-forms come actually from dimensionally reducing higher RR and NS fields which are related, at the classical level, by Hodge  duality. In fact, we have seen that the Hodge dualities relate the 4-form backgrounds and the internal fluxes forcing this extra shift to vanish. 
However we do not discard completely the possibility of an integer 
quantum shift which would not be visible at the level of the classical equations of motion here considered.

The underlying well-defined structure of the scalar potential in terms of the 4-forms is also remarkable. In this description it is clear that the solution of minimum energy will correspond to have all 4-forms vanishing, and can be obtained by solving eqs.(\ref{F4m}) and (\ref{H40}) in which the left side of each equation is equal to zero. We recover then the AdS minima with $b_i=q_i/m$ previously studied in detail in \cite{Luis2,dewolfe,moredual}. This suggests that moduli fixing might be more intuitive in terms of 4-forms.

Note that there are in total $2h_-^{11}$  4-forms, denoted above as $F_4^i$ and $F_4^a$,  which act as auxiliary fields for the  $h_-^{11}$ 
Kahler moduli of the compactification. This  means that the SUSY multiplets  associated to the Kahler moduli should contain two 4-forms
acting as auxiliary fields. On the other hand there are $h_+^3$ 4-forms $H_4^I$ associated to the $h_+^3$  complex structure fields. 
In this case the associated SUSY multiplets would only include one 4-form auxiliary field, like the multiplets discussed in
\cite{louis3form}.  In addition there are two  4-forms $F_4^0,F_4^m$ which seem to be associated to the 
$N=1$ supergravity complex scalar auxiliary field.
In this connection the relation imposed by the equations of motion between the 4-forms and the moduli of the compactification is interesting.
 By looking at eqs.(\ref{F4m}), the Minkowski 4-forms satisfy
\beqa
kF_4^0-v_aF_4^a=ReW\\
\frac12 k_iF_4^i-kF_4^m=ImW
\eeqa
where $W$ is the $N=1$ type IIA  RR superpotential given by
\beq
W=e_0+ie_a T^a-\frac12 k_{abc}q^aT^bT^c+\frac16 imk_{abc}T^aT^bT^c
\eeq
It would be interesting to understand if this structure is consequence of the possible identification of 4-form fields as auxiliary fields of the moduli/gravity multiplets. More generally, it would seem that non-minimal $N=1$ supergravity formulations, with auxiliary field scalars replaced by Minkowsk 4-forms, as in refs. \cite{gates1,gates2,binetruy,ovrut,binetruy2,girardi2,louis3form},  
could be the appropriate formulation to describe the  multi-branched nature of string flux vacua.

\subsection{Symmetries\label{sec:symmetries}}

The above effective action features  remarkable shift and duality symmetries which play an important role in
constraining the structure of the scalar potential.
In particular the latter is invariant under discrete  group transformations acting both on the moduli and the internal fluxes. They correspond to shifts on the axionic components of the Kahler and complex structure moduli combined with the corresponding changes on the internal fluxes. In particular, a shift on the Kahler axion given by
\beq
b_i\rightarrow b_i+n_i
\label{shiftb}
\eeq
combined with
\beqa
m  \rightarrow \rho_m&=&m\\
q_a \rightarrow  \rho_a(b_i=-n_i)&=&q_a+n_am\\
e_i   \rightarrow \rho_i(b_i=-n_i)&=& e_i-k_{ijk}q^j n^k-\frac{m}{2}k_{ijk}n^jn^k\\
e_0  \rightarrow \rho_0(b_i=-n_i)&=& e_0-e_i n_i +\frac12 k_{ijk}q^in^jn^k+\frac{m}{6}k_{ijk}n^in^jn^k
\label{transfis}
\eeqa
leaves invariant the scalar potential and relates equivalent vacua. 
These transformations were first introduced in the toroidal orientifold of ref.\cite{dewolfe}. They are however part of the duality symmetries
of any CY orientifold.  In the mirror  Type IIB picture this corresponds to a shift on the complex structure of the torus. Notice that the above transformations leave invariant each 4-form independently, as expected by coming from higher dimensional gauge invariance. Therefore the derivation of this group of transformations is more intuitive in this formulation in terms of 4-forms than in the standard Cremmer et al supergravity description. They also correspond to the generalization of the Kaloper-Sorbo shift symmetry underlying the axion monodromy inflationary models.

Analogously, the scalar potential is also invariant under shifts on the complex structure moduli of the form
\beqa
\label{shiftc}
c_3^I\rightarrow c_3^I+n^I\\
e_0\rightarrow e_0+h_I n_I
\eeqa
corresponding to the mirror of Type IIB SL(2,Z) shifts. Also in this case, the 4-forms remain invariant independently.

In a different vein, the effect of performing two or more T-dualities over the system is interesting .
Let us consider for simplicity a Type IIA toroidal orientifold compactification, and focus on the diagonal Kahler moduli.
The results can be generalised to other geometries with non-trivial one-cycles. 
Given a basis of 2-forms $\omega_i$ such that the Kahler form can be written as
\beq
J=\sum_{i=1}^3v^i \omega_i
\eeq	
we can perform two T-dualities along the two real directions of the Poincare-dual 2-cycle of some $\omega_i$. In particular, if T-duality is performed along $i=3$ we obtain again a type IIA theory in which
\beq
v^3\rightarrow \frac{1}{v^3}
\eeq
and the other  two fields $v^i$ with $i\neq 3$ remain invariant. 
In this case $v^3$ corresponds to the area of the 2-torus along which we perform the two T-dualities. Factors on $\alpha '$ are omitted to avoid clutter but can be easily recovered. Let us assume for simplicity an isotropic compactification such that the triple intersection number is $k_{ijk}=1$ if all the indices are different, and zero otherwise. The volume of the overal manifold transforms as
\beq
k=\frac16 k_{ijk}v^iv^jv^k=v^1v^2v^3\rightarrow \frac{v^1v^2}{v^3} 
\eeq
The metric is given in general by
\beq
g_{ij}=-\frac14 \left( \frac{k_{ij}}{k}-\frac14 \frac{k_ik_j}{k^2}\right)\ ,\quad g^{ij}=-4k\left(k^{ij}-\frac{v^iv^j}{2k}\right)
\eeq
and transforms under the two T-dualities as
\beq
\frac{g^{33}}{8k}\leftrightarrow \frac{1}{4k}\ ,\quad \frac{g^{11}}{8k}\leftrightarrow k g_{22}\ ,\quad \frac{g^{22}}{8k}\leftrightarrow k g_{11}\ ,\quad 2kg_{33}\leftrightarrow\frac{k}{2}
\label{dualidades}
\eeq
The RR part of the scalar potential 
is invariant under this T-duality if the functions defined in \eqref{rho} are also interchanged
\beqa
\rho_0\leftrightarrow \rho_i \quad\text{if }i=3\label{rho0}\\
\rho_i\leftrightarrow \rho_a \quad\text{if }i\neq a\neq 3\\
\rho_a\leftrightarrow \rho_m \quad\text{if }a=3\label{rhom}
\eeqa
Therefore T-duality seems to exchange Minkowski 4-forms with each other. Recall that each 4-form comes from dimensionally reducing the field strength of the different higher dimensional RR fields. Then it can be checked that the result matches with the known transformation rules for the RR fields under T-duality,
\beqa
C_3\leftrightarrow C_5 \quad &\text{if $C_5$ propagates along the T-dual direction}\\
C_5\leftrightarrow C_7 \quad &\text{if $C_7$ (but not $C_5$) propagates along the T-dual direction}\\
C_7\leftrightarrow C_9 \quad &\text{if $C_9$ (but not $C_7$) propagates along the T-dual direction} \ .
\eeqa

Finally, if the internal manifold is $T^6$ we can perform a T-dual transformation along all the internal dimensions, obtaining
\beq
k\leftrightarrow \frac{1}{k}\ ,\quad \frac{g^{ij}}{8k}\leftrightarrow kg_{ij}
\eeq
and the potential is invariant if
\beqa
\rho_0\leftrightarrow \rho_m \\
\rho_i\leftrightarrow \rho_a
\eeqa
consistent with the transformation rules for the RR fields.
Note that the fact that T-dualities relate the different 4-forms make that e.g. only the full
$V_{RR}$ combination, involving all 4-forms, will  be invariant under dualities and shift symmetries,
we will come back to this issue in section 5.  Let us conclude by mentioning that non-vanishing values for the
4-forms will generically break SUSY (since they are auxiliary fields). However the discrete symmetries will remain unbroken, since 
the 4-forms are invariant under them.

\subsection{4-forms and geometric fluxes in toroidal Type IIA orientifolds}

It is known that  beyond standard RR and NS other, less well studied NS fluxes 
may be present. These include the geometric fluxes in toroidal models that appear in the context of Scherk-Schwarz reductions.
In this section we will just explore whether the addition of these  fluxes change in any important way 
the above discussion.
 We will take an effective viewpoint on geometric fluxes and focus on the essential results. See  \cite{Villadoro,Luis2} and references therein
 for a more thorough discussion of geometric fluxes.

We are interested to see how the presence of geometric fluxes change the 4-forms described in eqs.(\ref{F4m}). Geometric fluxes are
 easiest described on a  factorized 6-torus $\otimes_{i=1}^3 T^2_i$ with O6-planes wrapping 3-cycles. In addition we assume there is a $Z_2\times Z_2$ orbifold twist
 so that only diagonal moduli survive projection. In this case we are left with 3 Kahler moduli and 4 complex structure moduli (including the 
 complex dilaton).
 In this setting there are $12$ geometric fluxes $\omega^M_{NK}$ that are convienently put in a 3-vector $a_i$ and a $3\times 3$-matrix $b_{ij}$,
 see \cite{Luis2,BOOK}  for  notation.

Geometric fluxes can be used to convert a $p$-form into a $(p+1)$-form via: $(dX)_{N_1\dots N_{p+1}}= \omega^K_{[N_1N_2}X_{N_3\dots N_{p+1}]K}$, denoted by $\omega \cdot X$. In particular we find:
\begin{align}
\omega\cdot B = b^ia_i \beta_0 - b^i b_{ij} \beta^j \quad \text{and} \quad \omega\cdot C_3 =-\tilde{\omega}^i a_i c_0 + \tilde{\omega}^i b_{ij} c^j \ .
\end{align}
From an effective viewpoint, geometric fluxes change the field strengths of $B$, $C_3$ and $C_5$ as follows \cite{Villadoro}:
\begin{align}
G_4 \rightarrow F_4 + \omega\cdot C_3 - H \wedge C_1 - \omega\cdot B_2 \wedge C_1 +\mathcal{F}e^B \ , \\
G_6 \rightarrow F_6  - H \wedge C_3 - \omega\cdot B_2 \wedge C_3  +\mathcal{F}e^B \ , \\
H_3 \rightarrow H_3 + \omega\cdot B_2 \ .
\end{align}
Putting these field strengths in the in the IIA action and integrating over the internal dimensions as before we find an extra coupling in the NS sector,
\begin{align}
-\int_{Y}e^{-2\phi} \omega \cdot B\wedge H_7 \quad = \quad  \frac{e^{-2\phi_4}}{k}\left( b^i a_i H^0_{4} -b^i b_{ij} H^j_{4}  \right) \ ,
\end{align}
and two in the RR sector,
\begin{align}
- \int_Y G_4\wedge G_6 \quad &= \quad  F^0_4 \left(b^i b_{ij} c^j -b^i a_i c0 \right)-F^i_4(b_{ij}c^j-a_ic^0) \ .
\end{align}
In this way the 4-forms get modified as
\begin{align*}
\star F_4^0 =  \frac{1}{k}[e_0+b^ie_ i-\frac{1}{6}mk_{ijk}b^ib^jb^k+\frac{1}{2}k_{ijk}q_ib^jb^k-h_0c_3^0-h_ic^i_3+b^ib_{ij}c^j_3-b^ia_ ic^0_3 ] \\
\star F^i_4= \frac{g^{ij}}{4k}[e_j+k_{jkl}b^kq^l-\frac{m}{2}k_{jkl}b^kb^l+b_{jk}c_3^k-a_jc^0_3] \\
\star H^0_4 =  h^0+b^ia_ i \\
\star H^i_4 =  h^i-	b^jb_{ji}  \ .
\end{align*}
The intersection numbers $k_{ijk}$ are equal to one  if all the indices are different and $0$ otherwise, since we have a toroidal compactification space. It can be shown that the scalar potential that is obtained from these 4-forms and eq.(\ref{VIIA}) can also be obtained from the superpotential given  in \cite{Villadoro,Luis2}. 

One interesting question  is how the discrete symmetries are modified in the presence of geometric fluxes.
One finds that the  4-forms are still  invariant under shifts of the  axion in the Kahler moduli and complex structure moduli
\begin{align}
b^i \rightarrow b^i+n_b^i &\, \\
c^J \rightarrow c^J+n_c^J &\,
\end{align}
in combination with
\begin{align}
h^0 \rightarrow h^0-a_in_b^i &\, \\
h^i \rightarrow h^i+n_b^jb_{ji} &\, \\
e_j \rightarrow e_j+a_jn_c^0-b_{jk}n_c^k&\, \\
e_0 \rightarrow e_0+h_in_c^i+h_0n^0_c+n_b^ib_{ij}n^j_c-n_b^ia_in_c^0 &\,
\end{align}
in combination with the shifts of the previous section. 
All in all, the general structure for 4-forms we described above remains in the pressence of geometric fluxes, as expected.

\section{4-forms in Type IIB  orientifolds }

We turn now to the case of Type IIB $D=4$ orientifolds. We concentrate on 4-forms coming from
the closed string sector but we also briefly mention an example of 4-form arising from the 
open string sector.

\subsection{4-forms and the IIB flux induced scalar potential}

Compared to Type IIA the structure in  IIB  \cite{Grimm:2004uq,Jockers:2004yj} is in principle 
slightly simpler because the CS couplings are simpler. Only the NS $H_3$ and RR $F_3$ tensors have a role
in the context of CY $N=1$  orientifolds.  It is convenient to define the complex 3-form
\beq
G_3\ =\ F_3\ -\ iSH_3
\eeq
where  $S$ is the complex dilaton, $S=1/g_s+ic_0$.  The relevant piece in our discussion are the 
kinetic terms of the 2-forms, which in this complex notation may be written as
\beq
\label{SIIB}
S_{IIB}=-\frac{1}{2k^2_{10}}\int_{\mathbb{R}^{1,3}\times Y} \frac{1}{3!} \frac{1}{S+S^ *}G_3\wedge ^*  {\bar G}_3 
\eeq
where $^*\bar G_3=\bar G_7$. As we did in the Type IIA case we can now expand $G_7$ in terms of internal harmonics with coefficients given by 
Minkowski 4-forms. We will consider here only IASD $G_3$ fluxes of class $(3,0)$ and $(2,1)$, which can induce SUSY-breaking. The contribution from ISD $G_3$ fluxes does not depend on the moduli and it is proportional to the topological number giving the  flux contribution to the D3 RR charge 
\cite{GKP,kst}, so it appears (combined with the contribution from localised sources) in the tadpole cancellation conditions. Then the relevant expansion is given by
\beq
{\overline G}_7 \ =\ {\overline G}_4^0\wedge \  {\overline \Omega}  \ + \ {\overline G}_4^a\wedge \chi_a  \ ,\ a=1,..,h_{21} \ ,
\eeq
where  $\Omega$ is the holomorphic $(3,0)$ form, and  the $\chi_a$ form a basis of the $h_{21}$ 3-forms in the
CY $X$. Here $G_4^0$ ans $G_4^a$ are complex Minkowski 4-forms which may be written in terms of 
NS and RR pieces $F_4,H_4$  as 
\beq
G_4^0 \ = \ F_4^0\ -iSH_4^0 \ ,\ G_4^a \ = \ F_4^a\ -iSH_4^a \ .
\eeq
The basis of $(2,1)$ forms may be expressed  in terms of the holomorphic 3-form
$\Omega$ and the complex structure Kahler potential $K$ as \cite{philip}
\beq
\chi_a \ = \  \frac {\partial \Omega}{\partial U_a} \ + \ K_{U_a}\Omega \ .
\eeq
From the term $G_3\wedge \bar G_7$ we then get  the kinetic terms
 for the Minkowski 3-forms and some couplings to the dilaton and complex structure moduli. For $G_4^a$ the coupling is given by
\beqa 
& \ & \frac {1}{S+S^*} \sum_a {\overline G}_4^a    \int_{X} G_3 \wedge
\left( \frac {\partial \Omega}{\partial U_a} \ + \ K_{U_a}\Omega \right)\ = \\ \nonumber
& = & \frac {1}{S+S^*} \sum_a {\overline G}_4^a \ D^a \int_X G_3 \wedge \Omega
\ = \ \frac {1}{S+S^*} \sum_a {\overline G}_4^a \ D^a W_{GVW}
\label{aform}
\ , 
\eeqa
where $W_{GVW}$ is the Gukov-Vafa-Witten superpotential and $D^a$ are the
Kahler covariant derivatives with respect to the complex structure fields $U_a$.   For the remaining 
4-form $G_4^0$ one gets the coupling
\beq
\frac {1}{S+S^*} {\overline G}_4^0 \int_X G_3 \wedge {\overline \Omega}
\ = \ -   {\overline G}_4^0 \     ({\overline {D_SW}}_{GVW}) \ .
\eeq
The kinetic terms of the 7-form yield
the quadratic pieces
\beq
\frac {\kappa }{S+S^*}(|G_4^0|^2  \ - \
G_4^a{\overline G}_4^b G_{a{\bar b}}) \ ,
\eeq
where $G_{a{\bar b}}$ is the metric of the complex structure fields and
\beq
\kappa = \int_X \Omega \wedge {\overline \Omega} \ =\ i e^{-K_{c.s.}(U_a)}  , \
G_{a{\bar b}}  \ = \  - \   \frac {\int_X    \chi_a \wedge  \overline{\chi}_b }{ \int_X \Omega \wedge {\overline \Omega} } \  ,
\eeq
where $K_{c.s.}(U_a)$ is the Kahler potential of the complex structure moduli. Collecting all the pieces, the ten dimensional action \eqref{SIIB} reduces to the following four dimensional effective lagrangian in terms of the Minkowski 4-forms,
\beq
\mathcal{L}_{IIB}=\frac {1 }{S+S^*}\left(\kappa\ (|G_4^0|^2  \ - \ G_4^a{\overline G}_4^b G_{a{\bar b}}) \ - {\overline G}_4^0 \  (S+S^*)   {\overline {D_SW}}_{GVW}+ \sum_a {\overline G}_4^a \ D^a W_{GVW}\right)
\label{LIIB}
\eeq
Notice that, in analogy to  Type IIA,  the full scalar potential, excluding the contribution from localised sources, can be written in terms of the Minkowski 3-form fields.
 One can now introduce Lagrange multipliers  enforcing $dC_3=F_4$ for each of the Minkowski 3-forms. Upon using the equations of motion for 
the 4-forms one gets
\beqa
\label{eqsIIB}
^*G_4^{\bar b}\  &=&\ -ie^{K_{c.s.}}G^{a{\bar b}}\left(D_aW_{GVW} \ + \ ( f_4 \ - \ iSh_4 )_a \right) \ , \\ \nonumber
^* G_4^0\  &=&\ -ie^{K_{c.s.}}\left((S+S^*) {\overline {D_SW}}_{GVW}\ + \ ( f_4 \ - \ iSh_4 )_0 \right)
\eeqa
where $f_4^{a,0}, h_4^{a,0}$ are RR and NS constants, from the Lagrange multipliers. We thus see that the complex 4-forms
$G_4^{a,0}$ are associated to the auxiliary fields of the complex structure and dilaton, but
include also a shift associated to the Minkowski 4-form backgrounds. This result is very similar to
the one discussed around eq.(\ref{shift1}) and suggested in \cite{binetruy,louis3form,louisdavid}.
The main difference is that here the two $N=1$ auxiliary fields are replaced by 4-forms and also
here it is the {\it supergravity}  auxiliary fields , with the covariant derivatives $D_a,D_S$, which 
suffer the shift.  Also the shift depends on the complex dilaton. In Type IIA we argued that the 
classical Hodge  dualities forced this shift to vanish, identifying the constant background terms of the Minkowski 4-forms with the internal fluxes of the magnetic duals. The analogy here would be to set $f_4,h_4=0$ with the argument that the internal fluxes parametrizing the $G_3$ background are enough to account for all the degrees of freedom (there should not be extra parameters). However we do not discard completely the possibility of an integer  quantum shift, not visible in the classical expressions
here considered.

By inserting eqs.\eqref{eqsIIB} in the Lagrangian \eqref{LIIB} we get the following scalar potential
\beq
V=e^{K_S+K_{c.s.}}\left( |(S+S^*)\overline{D_SW}+g_0|^2 +K^{a\bar b}|D_aW-g_a|^2\right)
\eeq
where we have used that $K_S=-log(S+S^*)$ and redefined $g_{0,a}\equiv ( f_4 \ - \ iSh_4 )_{0,a}$. If the shifts vanish, we recover the standard formula for the $N=1$ supergravity scalar potential.
Note that, due to the no-scale structure (there is no dependence of the superpotential on the
Kahler moduli), after using the equations of motion for the 4-forms one obtains a positive definite
scalar potential of the standard no-scale form. 

Finally, the same web of transformations studied in section \ref{sec:symmetries} relating different vacua of Type IIA compactified in a $CY$ threefold, are also present in Type IIB compactified in the mirror  $\tilde{CY}$. The discrete shift given by \eqref{shiftb} acting on a Kahler modulus of Type IIA corresponds to a shift on the complex structure of the mirror Type IIB. Notice that since the supergravity description to leading order in $\alpha'$ is reliable at large volume, this shift symmetry will arise at the large complex structure limit of the mirror $\tilde{CY}$. If we are dealing with a toroidal compactification instead, then the shift symmetry will correspond to the usual complex structure reparametrizations of the torus. Recall that this shift on the complex structure (in Type IIB) or in the Kahler modulus (in Type IIA) leaves invariant the effective theory only if it is combined with the corresponding transformations on the internal fluxes, studied in section \ref{sec:symmetries}. Analogously a shift on the complex structure \eqref{shiftc} in Type IIA maps to a shift on the axionic component of the Kahler moduli in Type IIB. This latter shift symmetry is expected from the fact that the imaginary part of the a Kahler modulus in IIB is actually an axion coming from dimensionally reducing the RR field $C_4$. 

While in Type IIA the current description in terms of 4-forms offered a very intuitive picture about these transformation (leaving each 4-form invariant), the situation in Type IIB is less transparent. Since we only have the 4-forms coming from $G_7$ we can not decompose the scalar potential into different smaller invariant pieces. Hence in the end the exercise of finding the transformation rules in this description is not easier than just studying the symmetries of the full scalar potential (or of the auxliliary fields). We would like to remark though that this set of transformations leave each 4-form invariant and are the generalization of the shift symmetry of axion monodromy models and the Kaloper-Sorbo Lagrangian. In other words, string theory provides a rich and more intriguing web of duality symmetries which are the generalization of the aforementioned shift symmetry. Besides, the full appearance of the `axionic' moduli in terms of couplings to the Minkowski 3-form fields highly constrains the form of $\alpha'$ and perturbative corrections. As we will see in section \ref{sec:stability} this structure acts as a sort of `chiral symmetry' protecting the scalar potential from dangerous higher order corrections, apparently independently whether the field appears or not in the 4d Kahler potential.

Before closing this section  let us make a few remarks about non-geometric fluxes \cite{non geometric,more dual} in toroidal 
Type IIB orientifolds,  see e.g.(\cite{BOOK}) for a brief description of these fluxes.
Non-geometric fluxes are still poorly understood although their existence is implied by T-dualities.
They are known to induce additional terms  in the superpotential. 
Type IIB orientifolds allow only for  so called $Q$-type non-geometric fluxes (we use notation in \cite{more dual})).
The fluxes have index structure $Q_M^{NP}$ with antisymmetric upper indices and they are odd 
under the $O(3)$ orientifold involution.   In IIB there are no geometric $\omega_{NP}^M$  nor 
non-geometric $R^{MNP}$ fluxes which are even.  The effect of the $Q$-fluxes on the 
Gukov-Vafa superpotential is captured by the replacement
\beq
G_3=(F_3-iSH_3) \ \longrightarrow G_3+{\cal Q J}_c
\eeq
where the 4-form
\beq
{\cal J}_c\ =\ i\sum_{i=1}^3 T_i{\tilde \omega }_i \ ,
\eeq
with $T_i$ the three diagonal Kahler moduli and
\beq
({\cal QJ}_c )_{MNP} \ =\ \frac {1}{2} { Q}_{[M}^{AB}({\cal J}_c)_{NP]AB} \ .
\eeq
Going back now to the 4-forms in IIB, eq.(\ref{aform}) gets modified as
\beq
\frac {1}{S+S^*} \sum_a {\overline G}_4^a \ D^a \int_X (G_3+{\cal Q_J}_c) \wedge \Omega \ .
\eeq
The right-hand side is nothing but the Kahler derivative (with respect to the complex structure) of the
extension of the GVW super potential to include non-geometric fluxes.
So it seems that also in the presence of this class of non-geometric fluxes
the structure of the Minkowski 4-forms acting as auxiliary fields in the
effective action persists.

\subsection{Minkowski 4-forms and open string moduli\label{sec:open}}

Up to here we have discussed the role of Minkowski 3-form fields in the closed string sector of Type II.  We have seen that the full RR and NSNS axion dependence of the flux scalar potential can be written in terms of these 3-forms. A similar question arises for the open string sector of the theory. Can also the scalar potential of the open 
(periodic) string moduli be written in terms of 3-form fields? In this section we address the issue for the D7-brane moduli sector of a Type IIB orientifold compactification. In particular, we review the computation done in \cite{higgsotic}, for which the flux induced scalar potential of a D7 position modulus can be written in terms of a Kaloper-Sorbo coupling of the scalar with a Minkowski 3-form field arising from the magnetic open string field strength. In \cite{higgsotic} the goal was to derive the Kaloper-Sorbo Lagrangian of a concrete inflationary model, dubbed Higgs-otic inflation, in which the inflaton is the position modulus of a wandering D7-brane in a transverse torus. This way one can use the Kaloper-Sorbo symmetry properties to argue from an effective approach that the higher order corrections are under control and do not spoil inflation. Here we derive the effective theory and discuss the result with the new insight gained from previous sections.

In the open string sector of Type II string theory, Minkowski 3-forms may arise from the dual magnetic potentials of the worldvolume gauge fields of the D-branes. In particular, for a D7-brane the magnetic gauge potential is a 5-form $A_5$, whose field strength can be expanded as
\beq
F_6=iF_4\wedge \bar\omega_2-i\bar F_4\wedge \omega_2
\label{F6}
\eeq
where $\omega_2$ is a (2,0)-form associated to the position modulus $\Phi$ of the D7. This field can be expanded as $\Phi=\phi\omega_2$ where $\phi$ is a 4d complex scalar. Notice that unlike the 4-forms coming from the closed string sector, now $F_4$ is a complex Minkowski 4-form.  We are going to focus on the Abelian case, but a priori it could be generalised to non-Abelian gauge groups.

Consider ISD $G_3$ bulk fluxes inducing a B-field on the brane given by  \cite{higgsotic,civ,Camara:2003ku,Grana:2003ek,Lust:2004fi,Camara:2004jj}
\beq
B_2=\frac{g_s\sigma}{2i}(G^*\phi-S\bar \phi)\omega_2+cc. \ \ ,
\label{B2}
\eeq
where we have denoted the non-supersymmetric  ISD (0,3)-flux as $G\equiv G_{\bar 1\bar 2\bar 3}$ and the supersymmetric (2,1)-flux as $S\equiv \epsilon_{3jk}G_{3\bar j\bar k}$
(see \cite{Camara:2004jj} for notation).
The relevant part of the DBI action to leading order in $\alpha'$, ie. in the Yang-Mills approximation, is given by \cite{Camara:2004jj,civ,higgsotic}
\beq
\mathcal{S}_{DBI}=\mu_7\sigma\int_{\mathbb{R}^{1,3}\times S_4} \ \frac12(B_2+\sigma F_2)\wedge *_8  (B_2+\sigma F_2)+\dots
\eeq
where $\sigma=2\pi\alpha'$. Plugging the decomposition \eqref{F6} into the above Lagrangian and performing dimensional reduction we obtain
\beq
\int_{S_4} \ F_6\wedge *_8F_6=|F_4|^2\ 2\int_{S_4}\omega_2\wedge *_4\bar \omega_2 \ ,
\label{F2F6}
\eeq
\beq
\int_{S_4}\ B_2\wedge F_6= \frac12 g_s\sigma \left(F_4(G^*\phi-S\phi^*)+\bar F_4(G\phi^*-S^*\phi)\right)\int_{S_4}\omega_2\wedge\bar\omega_2 
\ ,
\eeq
leading to the following effective four dimensional Lagrangian
\beq
\mathcal{L}_4=\mu_7\sigma\rho \ \left(|F_4|^2-\frac12 g_s\sigma \left(F_4(G^*\phi-S\phi^*)+\bar F_4(G\phi^*-S^*\phi)\right)\right)+\dots
\eeq
Here  $\rho=\int_{S_4}\omega_2\wedge *_4\bar \omega_2$ and we have used that $*_4\omega_2=-\omega_2$. Upon integrating out the 3-form field we get 
\beq
V_4=\mu_7\sigma\rho \ \left|f-\frac12 g_s\sigma(G^*\phi-S\phi^*)\right|^2 \ ,
\label{higgsoticbis}
\eeq
where $f$ is an integration constant which can be identified with the magnetic flux $F_2$.
Note however that, by an appropriate choice for the $B_2$ gauge, the constant term $f$ may be reabsorbed into the definition 
of what the origin of the wandering D7-brane is. In fact it was took equal to zero in \cite{higgsotic}. 
The above expression reflects the branched properties of the Higgs vev as the D7-moves along a cycle in the torus.
The potential is invariant under shifts on the position modulus if they are combined with the corresponding shift on $F_2$ flux. This shift symmetry underlies the typical multi-branch
structure of a Kaloper-Sorbo Lagrangian (or an F-term axion monodromy model), its presence being  important to keep the potential under control in large field inflationary models. The idea again is that the underlying shift symmetry and the gauge invariance of the 3-form field protects  the potential from dangerous higher order corrections, as we will discuss in section \ref{sec:stability}. Once a specific branch is chosen, ie. the flux background is fixed, we can inflate with the position modulus inducing the monodromy and allowing for large field excursions.
Let us finaly note  that we recover only half of the complete scalar potential because we are missing the Chern-Simons part of the action, which because of supersymmetry will give the same contribution as in eq.\eqref{higgsoticbis} (see \cite{Grimm:2008dq} for a similar computation of the flux-induced scalar potential obtained from the Chern-Simons action of a D5-brane, in which the authors also keep explicetely the presence of the 4-forms coming from the RR fields).

In addition to the above quadratic piece, the position modulus can also have couplings with other matter fields. Yukawa couplings with the D7 Wilson lines $A_I$ can be obtained by considering the non-abelian part of $F_2$ in eq.\eqref{F2F6} and can also be written in terms of the Minkowski 3-form field.

One can think of exploring a similar structure within Type IIA orientifolds.
The magnetic gauge field is a 4-form $A_4$, which has to be expanded in a basis of 1-forms on the $D6$-brane 3-cycles
 in order to get a Minkowski 3-form field.
 This can be done e.g.  in toroidal models, recovering the T-dual picture of the D7-brane models discussed above, and  also CY's with appropriate
 3-cycle topology.  One can  expand also  the magnetic gauge field in a basis of torsion cycles obtaining the models of Massive Wilson lines discussed in \cite{msu}.
 We leave the study of this IIA case for future work.


\section{4-forms, inflation and stability of scalar potentials\label{sec:stability}}

\subsection{ Stability of axions in flux vacua}

In this section we discuss possible physical  consequences of the structure of flux vacua described in terms of
minkowski 4-form fluxes as discussed above.  For these applications a crucial point  is that we have found that
all the RR and NS axion dependence of the flux scalar potential goes always through Minkowski 3-forms.  And by
gauge invariance of the latter, the flux potential, even after $\alpha'$ and perturbative corrections are considered, 
should admit an expansion in powers 
of the gauge invariant Minkowski 4-forms, i.e. for $V_{RR}$  in Type IIA 
\beq 
V(b_i,c_a) \ =\ \sum_{r,s_i,t_a,u}  \frac {c_{rs_it_au}}{m_p^{4(w-1)}}  \  (F_0^{2r})(F_m^{2u} )
(\Pi_{i}   ( g_{kl}F^kF^l)^{s_i})
(\Pi_{a}   ( g_{bc}F^bF^c)^{t_a})
   \ - \  V_{non-axionic}  \ ,
\label{sumasuma}
\eeq
where $r,u,s_i,t_a$ are integers and the $c$'s are coefficients depending on the non-axionic components of
the Kahler and complex structure moduli. $F_0,F_i,F_a,F_m$ are the 4-forms discussed in section 3 (contracted 
with the Levi-Civita tensor), and
\beq
w=r+u+\sum s_i +\sum t_a -4  \ .
\eeq
$V_{non-axionic}$ collects pieces of the potential which will not depend on the axion fields,
like the local contributions in $V_{local}$ discussed in section 3. Note that  for a single factor, e.g  $t_1\not=0$, $g_{11}=1$  with the rest of the integers vanishing,
one gets the familiar Kaloper-Sorbo structure  of the type
\beq 
V(b_1) \ =\   \sum_{n\geq 1} \frac {c_n}{m_p^{4(n-1)} } V_0^n\ =\  \sum_{n\geq 1} \frac {c_n}{m_p^{4(n-1)} } (F_1)^{2n} \ =\ \sum_{n\geq 1} \frac {c_n}{m_p^{4(n-1)} } (q_1\ + mb_1)^{2n} \ , 
\eeq
and a discrete symmetry  $q_1\rightarrow q_1-m$, $b_1\rightarrow b_1+1$.
As claimed in \cite{KS}, due to this symmetry and the gauge invariance of $F_1$, all corrections to the leading quadratic potential are  suppressed.
Indeed, if applied to inflation,  the axion/inflaton  $b_1$ can have large field trans-Planckian excursions  since, with a Hubble parameter at inflation $H_I\simeq 10^{16}$ GeV,
the possible corrections will be suppressed by powers $V_0/m_p^4$ and there will not be isolated $b_1^n/m_p^n$  terms in the potential.

As we have seen, in string theory the story is slightly more complicated, there are many 4-forms in the game and a  complicated
moduli fixing potential. Still the message we have found is similar. All axion dependence comes through 4-forms, which are gauge invariant,
and each 4-form is invariant under discrete transformations  under which the axions shift while internal fluxes also shift.  These transformations are a subset of the duality symmetries present in a given CY compactification. The axions are not 
real axions,  in the sense that they may have masses and Yukawa couplings.  They could be called {\it multi-branched  axions} since  they feature a discrete shift symmetry as long as 
internal fluxes are also shifted. This is the branched structure which has appeared in the past in the context of F-term monodromy inflation\cite{msu}. Moreover, the quadratic potential of Kaloper-Sorbo is replaced by more general polynomials up to order six.

In fact the situation in string theory is often much simpler than what eq.(\ref{sumasuma}) seems to indicate. Indeed, as we have shown in the Type IIA toroidal
orbifold example, although the 4-forms are invariant under axion shifts, they transform into each other under duality transformations, T-dualities in this example.
Due to the transformations in eq. (\ref{dualidades}) the different RR 4-forms appear in the particular combination $V_{RR}$ in eq. \eqref{VIIA} so that actually the corrections to the RR potential 
will appear as an expansion in powers of $V_{RR}$ itself. More generally, the full duality group of a specific Type II orientifold would often force the 
corrections to the original potential to be an expansion in powers of the leading potential  potential $V_0$, i.e.one expects  in these cases
\beq
V\ =\ \sum_n \ c_n V_0^n \ ,
\label{scalarchirality}
\eeq
where $V_0$ is the tree level, leading order in $\alpha '$,  flux potential.

In the applications of this setting to inflation, the higher order corrections in $V_0$, although under control, may be non-negligible in the
particular case of large field inflation. In fact in that case they may lead to a flattening effect \cite{flattenings} so that, although for small field
the uncorrected $V_0$ gives an appropriate description, the asymptotic behaviour  of the potential for large field  gets
flattened. This has been observed in the context of certain monodromy inflation models. It  may also appear as an effect of the
interaction of the inflaton with heavy modes.  This flattening effect is important, since e.g.  it makes standard quadratic chaotic inflation
become e.g. linear and be consistent with Planck/BICEP limits. Note however that this flattening effect does not modify the
value of masses at the origin.

In models of inflation, in which the inflaton appears in the Dirac-Born-Infeld action of a D-brane, the $\alpha'$ corrections to the 
scalar potential do appear as a series expansion in the leading scalar potential $V_0$, see e.g. \cite{mono,gurari,msu,higgsotic}, in agreement with \eqref{scalarchirality}.
It arises upon expansion of the square root in the DBI action or, in the case in which the scalar is an open string mode, due
to a non-canonical redefinition of the scalar kinetic term. An example of this effect is discussed in \cite{higgsotic} in which the 
inflaton (which in this case is identified with a MSSM Higgs) has an action with terms of the general form
\beq
{\cal L}\ =\ -(1+\xi V_0)(D_\mu \phi)^2 \ -\ V_0\ +... \ ,
\label{higgsotic}
\eeq
where in this case $V_0$ is just quadratic and $\xi$ is a constant factor proportional to $\alpha'$.
After setting the kinetic term in canonical form,  $\alpha'$ corrections to the potential appear 
as a power series in $V_0$, giving rise to a linear behaviour for large $\phi$. In this case 
$\phi$ parametrizes the position of a D-brane on a torus, which is T-dual to a continuous Wilson line.
Although here $\phi$ is not a closed string monodromy axion, the model is an example of monodromy inflation 
since there is a  shift symmetry corresponding to discrete translation of the D-brane of the torus and a 
non-trivial scalar potential arising from fluxes. As discussed in section \ref{sec:open} the scalar potential $V_0$ admits a description 
in terms of a complex open string Minkowski 4-form.

Let us also note in closing that kinetic term redefinitions like that appearing in eq.(\ref{higgsotic}) also appear in
computing the higher derivative corrections to general $N=1$ supergravity Lagrangians, see \cite{Aoki:2015eba,louiswestphal}
and references therein.

\subsection{Multi-branched axions, scalar stability  and naturalness}

There are essentially a two well stablished ideas in order to  make stable scalar masses against loop 
corrections and get naturally light scalars, i.e. naturalness. 
One of them is supersymmetry and the other is Goldstone bosons. To the latter case belong 
the (continuous) Peccei-Quinn shift symmetry of axions. This symmetry is only broken at the non-perturbative level by
instanton effects.  It has however the drawback that axions have only derivative couplings (e.g., no Yukawas).  Analogously, there are
BSM models in which the Higgs fields are Goldstones bosons of a spontaneously broken global symmetry but again 
Yukawa couplings are forbidden to leading order.

The class of flux potentials discussed in this paper seem to show the existence of a third alternative to achieve
scalar masses stable against loop corrections.   These are fields analogous to the {\it multi-branched  axions} discussed in the
previous sections.  These fields present a discrete shift  invariance when accompanied by adequate 
transformations of de Lagrangian parameters, fluxes in the case at hand. So it is important to realise that these {\it are not symmetries
of the field action}. Rather, this is a symmetry of a Landscape of Lagrangians differing by the different values of the 
parameters, fluxes in the string theory case.  But this Landscape structure, with different values for fluxes,   is what one really
finds in string theory. Membrane domain walls, coupled to 3-forms are able to
interpolate between potentials  corresponding to different Lagrangian parameters (fluxes).

A very important difference with the axion or,  in general,  Goldstone boson symmetry is that 
the symmetry is {\it consistent with interactions}. Consider as an explicit model that provided by the RR potential $V_{RR}$ of
Type IIA string compactifications in section 3.  The potential presents polynomial interactions of the scalars $b_i$, which are still 
consistent with  shifts $b_i\rightarrow b_i+n_i$ as long as the $e_0,e_i,q_a$ parameters are transformed as in eq.(\ref{transfis}). 
There is in fact a landscape of potentials corresponding to different choices for these parameters, and the symmetries relate 
different potentials and vacua.  In view of these symmetries and the landscape structure,  the
perturbative corrections to this scalar potential should appear as a power series in the tree potential $V_{RR}$. One expects for the
corrected potential to a be a function  $V=f(V_{RR})$ of the classical potential.
Assume that the uncorrected potential $V_{RR}^0(\phi_i)$ has a minimum at some values 
$\phi_i=\phi_i^0$, Then at that minimum
\beq
\frac {dV_{cor}}{d\phi_i} \ =\  (\frac {df}{dV_{RR}}) (\frac {dV_{RR}}{d\phi_i}) \ =\ 0
\eeq
so that the corrected potential has also in general a extremum there.  For the masses at that point one then has  
\beq
\frac {\partial^2V_{cor}}{\partial \phi_i\partial \phi_j} \ = \  
(\frac {\partial f}{\partial V_{RR}})(\frac {\partial^2 V_{RR}}{\partial \phi_i\partial \phi_j}) \ 
\eeq
and for $\frac {\partial f}{\partial V_{RR}}>0$ the corrected potential has also a minimum there.  

The obvious observation is that   both mass matrices are proportional and hence, for an analytic function $f$, if the uncorrected 
potential has a massless scalar, the corrected potential will also have a massles scalar. There exists  a sort of
{\it scalar chirality}  for the scalar potential if the structure (\ref{scalarchirality}) is  true.  In particular it would seem that modifications like e.g. a
quadratic loop divergence for the masses of scalars should be forbidden, since they would violate explicitly this property.
One way to understand this is to think that, from the effective field theory point of view, upon renormalisation we should have
to use a regulator which is consistent with  4-form gauge  invariance and the shift and duality symmetries, and that such a regulator should not allow for
a quadratic cut-off. While working on a given branch,  a perturbative calculation would yield e.g. a quadratic scalar divergence for  interacting scalars like these. 
However that would violate the branched structure of the theory. Once taking the latter  into account such divergences should be 
forbidden by the symmetries.

It is tempting to speculate about the application of this stability property to the  SM hierarchy problem and the Higgs. 
Although the Higgs field is not directly an axion, there are string constructions in which the Higgs degrees of
freedom may be identified with complex Wilson lines (see e.g. \cite{higgsotic} and references there in) or their T-dual,
D-brane position moduli. In such a case the scalar Higgs may posses  a multi-branch  structure and one could conceive such
an scenario. The symmetries would forbid quadratic divergences for the Higgs mass.  Still the scalar potentials considered in this paper
contain only moduli and no gauge interactions with charged fields. Furthermore, in order to be useful, one should be able obtain
a SUSY breaking scale much larger than the discrete symmetry breaking scale, so that it is the latter which is mantling scalars light
rather than SUSY. In this respect one should play with the different flux degrees of freedom and compact volumes, which requires
a complete scheme of moduli fix ing in De Sitter.
It would be interesting to see whether models with these
characteristics and the property (\ref{scalarchirality}) can be built.
We leave the study of that possibility to future research.

\section{Conclussions and outlook}

In this paper we have studied the role of Minkowski 3-forms in (orientifold) flux compactifications of Type IIA and Type IIB theory.
To this aim we have performed  an explicit dimensional reduction of the D=10 Type II actions  in the presence of
RR and NS internal fluxes, keeping trace of the resulting Minkowski
4-forms and their couplings.  These external fluxes are in one to one correspondance with the more familiar  internal fluxes.
We find that the Minkowski 4-forms act as  auxiliary fields of the Kahler and complex structure moduli of the
compactifications. This is consistent with the fact that 3-forms in Minkowski have no propagating degrees of freedom, 
but the corresponding field strength 4-forms may contribute to the cosmological constant.

We find that the   dependence of the flux scalar potential on the  RR and NS axions always goes through  the 
corresponding Minkowski 4-forms.  In this context is then important to realise that in any 
Type II orientifold  compactifications  there are  symmetries under which these  RR and NS axions  suffer discrete
shifts as long as appropriate discrete translations of the internal fluxes are performed. Interestingly, the 4-forms 
are invariant under these transformations. This,  combined with the (3-form) gauge invariance,  forces the 
axion scalar potentials, even after perturbative corrections are included,  to be expressible as an expansion in powers 
of the 4-forms. We also argue, and exemplify  in a Type IIA toroidal case, that additional duality symmetries will typically
force a more restrictive structure with a  perturbatively corrected scalar potential being a power series of the leading order 
potential $V_0$, as in eq.(\ref{scalarchirality}).  This would be  both a multifield generalisation with higher order couplings and a string realisation of the
field theory idea of Kaloper and Sorbo. An important difference though is that the axions in this paper posses non-trivial polynomial interactions.

The use of 4-forms, acting as auxiliary fields,  to describe  string flux vacua is most appropriate to reveal the multi branched structure 
of flux scalar potentials. We have found that in both Type IIA and Type IIB orientifolds  keeping track of the 4-forms appearing upon compactification 
allow us  to identify in a simpler way the underlying symmetries  of the flux multi branched vacua. The corresponding 3-forms couple to 
membranes which can break these symmetries but only at the non-perturbative level, inducing vacuum transitions through domain walls.

The discrete symmetries which are preserved by the 4-forms are not standard single Lagrangian symmetries but rather symmetries of a landscape of
Lagrangians parametrized by the different internal fluxes.  Not only fields transform, but masses and couplings (fluxes) as well.
The RR and NS fields are not axions in the usual sense (since their flux couplings 
break the Peccei-Quinn symmetry explicitly)  but {\it multi-branched axions} , which are only invariant under discrete symmetries accompanied by 
flux transformations. 
 The  property in eq.(\ref{scalarchirality}),  preserved in an $\alpha'$ expansion,   would be of interest for  large field models of inflation in the string theory context, 
 particularly in models of F-term monodromy \cite{msu}, which directly contain this Kaloper-Sorbo structure. The symmetries will protect the corresponding axion/inflaton 
 when featuring trans-Planckian trips.
 Furthermore one  expects  the loop corrections to preserve also the structure in (\ref{scalarchirality}).
In particular we have argued that quadratic divergences are not expected to appear for these {\it multi-branched axions} even though they can have
non-trivial couplings and masses and no supersymmetry.  
This would provide for a new mechanism to maintain interacting scalars stable against
quadratic loop corrections.  We hope to report on possible applications of these  ideas in future work.

\vspace{3cm}

\centerline{\bf \large Acknowledgments}

\bigskip

\noindent We thank G. Aldazabal, R. Kallosh,   F. Marchesano, M. Montero, A.~Uranga and G. Zoccarato for useful discussions. 
This work has been supported by the ERC Advanced Grant SPLE under contract ERC-2012-ADG-20120216-320421, by the grant FPA2012-32828 from the MINECO,  and the grant SEV-2012-0249 of the ``Centro de Excelencia Severo Ochoa" Programme.  I.V. is supported through the FPU grant AP-2012-2690 and S. B.  by the ERC grant SPLE.

\end{document}